\documentclass{aa}
\usepackage{graphicx}

\newcommand{\hii}{H~{\small II} }
\newcommand{\bra}{Br$\alpha$}
\newcommand{\brb}{Br$\beta$}
\newcommand{\bracket}{Br$\alpha$/Br$\beta$}
\newcommand{\neii}{[Ne~{\small II}]12.8$\mu$m}
\newcommand{\neiii}{[Ne~{\small III}]15.6$\mu$m}

\newcommand{\arii}{[Ar~{\small II}]6.99$\mu$m}
\newcommand{\ariii}{[Ar~{\small III}]8.99$\mu$m}
\newcommand{\siii}{[S~{\small III}]18.7$\mu$m}
\newcommand{\siiib}{[S~{\small III}]33.5$\mu$m}

\begin{document}

\headnote{Research Note}
\title{The ISO\thanks{Based on observations with ISO, an ESA project with
instruments funded by ESA Member States (especially the PI countries: France,
Germany, the Netherlands, and the United Kingdom) with the participation of
ISAS and NASA.} Galactic Metallicity Gradient Revisited}

\author{Uriel Giveon\inst{1}
          \and
          Christophe Morisset\inst{2}
          \and
          Amiel Sternberg\inst{1}
}

\institute{School of Physics and Astronomy and the Wise Observatory,
        The Beverly and Raymond Sackler Faculty of Exact Sciences,
        Tel Aviv University, Tel Aviv 69978, Israel \\
         \and
        Institut d'Astrophysique de Marseille, CNRS \& Univ. de Provence, BP 8,
	F-13376 Marseille Cedex 12, France\\ }

\date{Received ; accepted }
\titlerunning{Galactic Metallicity Gradient}
\authorrunning{Giveon et al.}

\abstract{
Two independent groups 
(Giveon et al. \cite{giveon}; Mart\'{\i}n-Hern\'{a}ndez et al. \cite{martin})
have recently investigated the Galactic metallicty gradient as
probed by ISO observations of mid-infrared emission-lines from HII regions.
We show that the different gradients inferred by the two groups are due to
differing source selection and differing extinction corrections. We show that
both data sets in fact provide consistent results if identical assumptions are
made in the analysis. We present a consistent set of gradients in which we
account for extinction and variation in electron temperature across the
Galactic disk.
}

\maketitle

\section{Introduction}

In two recent studies Giveon et al. (\cite{giveon}) and
Mart\'{\i}n-Hern\'{a}ndez et al. (\cite{martin}) analyzed mid-infrared
fine-structure emission lines, as observed by the {\it Infrared Space Observatory}
(ISO), to study the excitation and metallicity of HII regions across the
Galactic disk. Discrepancies appear to be present in the gradients
published in these two papers. In this note we re-analyze the reduced data
compiled by Giveon
et al. and  Martin-Hernandez et al. and we demonstrate that similar results
are obtained if the same source samples are chosen, and no assumptions are made
for the extinction correction. We then apply 
extinction and electron temperature
corrections to both data sets and infer consistent Galactic abundance gradients
from the two independent ISO studies.

\section{The Galactic Metallicity Gradient}
\label{gradient}

Giveon et al. (\cite{giveon}) presented a sample of 112
Galactic \hii regions, observed by ISO-SWS, spanning galactocentric radii 
$<$ 18 kpc. Most of their sources show 
prominent \arii, \ariii, \neii, \neiii, \siii\, and 
\siiib~fine-structure lines and 
also the hydrogen recombination lines \bra\ and \brb
at 4.05 and 2.60 $\mu$m.
Only 48
sources have both argon or neon lines and a physical Brackett ratio.
For case-B recombination, the \bracket\ ratio ranges from 1.7 to 1.9 for
temperatures $5\times10^3-2\times10^4$ K. Since extinction should only increase
the ratio, five sources with Brackett ratios
 $<1.68$ were excluded from the analysis. 
The Brackett line ratios, an assumed gas temperature of $10^4$ K,
and the Draine (\cite{draine})
mid-infrared extinction curve (see below)
were used to de-redden the fine-structure lines
intensities.

A similar analysis was performed by Mart\'{\i}n-Hern\'{a}ndez et al.
(\cite{martin}), who used the ISO-SWS \& LWS spectral catalog of 34 compact
\hii regions (Peeters et al. \cite{peeters}). The line intensities in their
paper were not corrected for dust extinction due to the small number of sources
for which the Brackett recombination lines were observed.
However, corrections due to possible electron electron temperature
variations across the disk were applied. In contrast, Giveon et al. did not
apply such corrections to their data.
Mart\'{\i}n-Hern\'{a}ndez et al. also published gradients of S/H and N/O.
Here we discuss only the common results as analyzed by 
Mart\'{\i}n-Hern\'{a}ndez et al. Giveon et al., and show
results for Ar/H, Ne/H, and S/H.

The results published in Giveon et al. and Mart\'{\i}n-Hern\'{a}ndez et al. are
listed in the first row of each panel in Table \ref{gradlist}.
\begin{table*}[]
\begin{center}
\caption{Galactic abundance gradients.}
\vspace{0.5cm}
\begin{tabular}[h]{lcccc} \hline
 & \multicolumn{4}{c}{NEON} \\ \hline
 & \multicolumn{2}{c}{Mart\'{\i}n-Hern\'{a}ndez Catalog} & \multicolumn{2}{c}{Giveon Catalog} \\
 ASSUMPTIONS & A$\pm\Delta A$ & B$\pm\Delta B$ & A$\pm\Delta A$ & B$\pm\Delta B$ \\ \hline
1. Published (2002$^1$) & $-3.49\pm0.06$ & $-0.039\pm0.007$ & $-3.71\pm0.06$ & $-0.021\pm0.007$\\
2. Sub-catalog, not corrected & $-3.58\pm0.09$ & $-0.024\pm0.011$ & $-3.67\pm0.08$ & $-0.03\pm0.01$ \\
3. Full catalog, not corrected          & $-3.49\pm0.06$ & $-0.036\pm0.007$ & $-3.61\pm0.06$ & $-0.019\pm0.008$ \\
4. Full catalog, att. corrected         & $-3.58\pm0.09$ & $-0.028\pm0.011$ & $-3.66\pm0.07$ & $-0.017\pm0.009$ \\
5. Full catalog, att.\& temp. corrected & $-3.48\pm0.09$ & $-0.044\pm0.011$ & $-3.55\pm0.06$ & $-0.035\pm0.009$ \\
\hline
 & \multicolumn{4}{c}{ARGON} \\ \hline
 & \multicolumn{2}{c}{Mart\'{\i}n-Hern\'{a}ndez Catalog} & \multicolumn{2}{c}{Giveon Catalog} \\
 ASSUMPTIONS & A$\pm\Delta A$ & B$\pm\Delta B$ & A$\pm\Delta A$ & B$\pm\Delta B$ \\ \hline
1. Published (2002$^1$)     & $-5.10\pm0.09$ & $-0.045\pm0.011$ & $-5.31\pm0.06$ & $-0.018\pm0.008$\\
2. Sub-catalog, not corrected & $-5.17\pm0.12$ & $-0.036\pm0.014$ & $-5.23\pm0.09$ & $-0.05\pm0.01$ \\
3. Full catalog, not corrected          & $-5.11\pm0.07$ & $-0.041\pm0.009$ & $-5.14\pm0.06$ & $-0.043\pm0.009$ \\
4. Full catalog, att. corrected         & $-5.20\pm0.09$ & $-0.030\pm0.012$ & $-5.25\pm0.06$ & $-0.026\pm0.008$ \\
5. Full catalog, att.\& temp. corrected & $-5.09\pm0.09$ & $-0.048\pm0.012$ & $-5.13\pm0.06$ & $-0.046\pm0.009$ \\
\hline
 & \multicolumn{4}{c}{SULFUR} \\ \hline
 & \multicolumn{2}{c}{Mart\'{\i}n-Hern\'{a}ndez Catalog} & \multicolumn{2}{c}{Giveon Catalog} \\
 ASSUMPTIONS & A$\pm\Delta A$ & B$\pm\Delta B$ & A$\pm\Delta A$ & B$\pm\Delta B$ \\ \hline
3. Full catalog, not corrected          & $-4.85\pm0.15$ & $-0.024\pm0.017$ & $-5.00\pm0.15$ & $-0.002\pm0.02$ \\
4. Full catalog, att. corrected         & $-4.95\pm0.15$ & $-0.015\pm0.020$ & $-5.01\pm0.15$ & $-0.001\pm0.02$ \\
5. Full catalog, att.\& temp. corrected & $-4.85\pm0.15$ & $-0.030\pm0.020$ & $-4.91\pm0.15$ & $-0.016\pm0.02$ \\

\hline
\end{tabular}
\label{gradlist}
\note{These rows show the results as originally published by
Mart\'{\i}n-Hern\'{a}ndez et al. (\cite{martin}) and Giveon et al.
(\cite{giveon}).}
\end{center}
\end{table*}
In this table we also list the argon, neon, and sulfur
abundance gradients determined
using a linear fit to the logarithmic abundances values, under various
assumptions, that we discuss below. The gradients coefficients are given in
the form A$\pm\Delta A$ and B$\pm\Delta B$, where
\begin{equation}
\left[{\rm X\over H}\right]=
A(\pm\Delta A)+B(\pm\Delta B)R_{\rm gal}({\rm kpc}),
\label{grad}
\end{equation}
where $X$ is argon, neon or sulfur.

If only the common sources for the two samples, and
no extinction corrections are applied, 
the two data-sets give comparable metallicity gradients. 
There are 22 sources in Giveon et al. which are common with
Mart\'{\i}n-Hern\'{a}ndez et al. We exclude the most distant source 
(WB89 380; IRAS 01045+6506)
in the samples, since the 
assumed distance to this source differs between the two
studies. For the 21 common sources we obtain the coefficients listed in
the second row of Table \ref{gradlist}. The results derived
from the common sources are in reasonable
agreement.
The small discrepancies may be due to differing 
($<5\%$) inferred line fluxes in the 
independent data reduction and small differences 
($< 10\%$) in the assumed galactocentric
positions of the sources.

\section{Combined Extinction and Temperature Corrections}
\label{results}

We now consider the gradients resulting from the full samples of Giveon et al.
(\cite{giveon}) and Mart\'{\i}n-Hern\'{a}ndez et al. (\cite{martin}), with 
corrections due to extinction and electron temperature dependence of the line
emissivities.

\begin{figure}
\centering
\includegraphics[width=9cm]{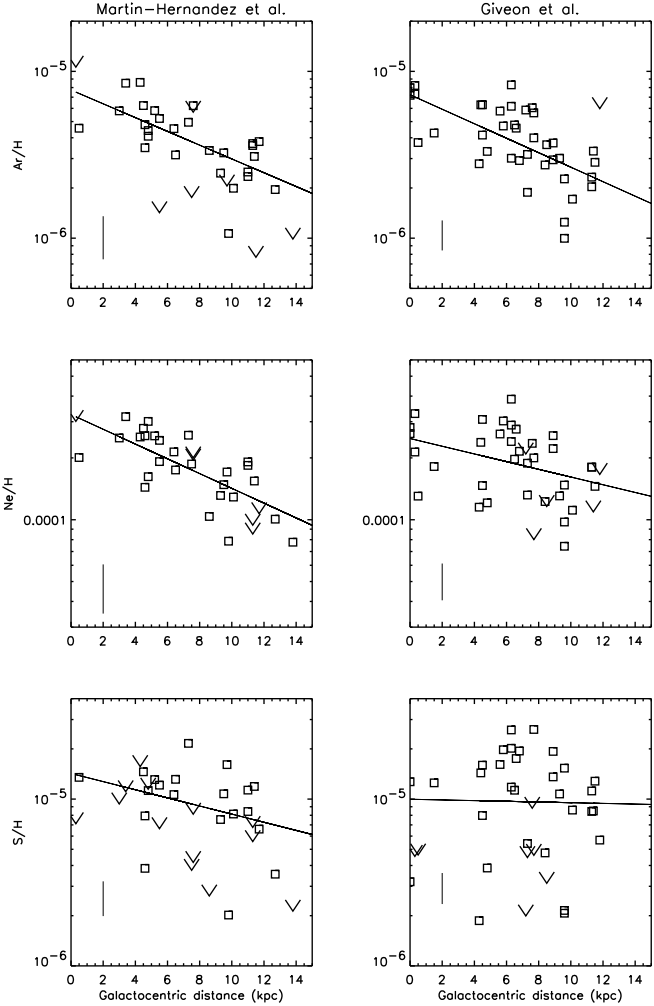}
\caption{Argon, neon, and sulfur abundances {\it vs.} galactocentric radius for the
full catalogs of Mart\'{\i}n-Hern\'{a}ndez et al. and Giveon et al. (raw data).
Solid lines are fits to the data, whose values are given in the third row of
Table \ref{gradlist}. In these plots no extinction or electron temperature
corrections have been applied. 
The vertical lines are representative error bars.}
\label{raw}
\end{figure}

We first take the entire samples and compute the abundance gradients
with no corrections applied. 
An electron temperature of $10^4$ K is assumed.
The resulting gradients are given in the third row of Table \ref{gradlist}.
The upper limits were included in these fits.
We show the total gradients for the uncorrected data in Fig.~\ref{raw}.

\begin{figure}
\centering
\includegraphics[width=9cm]{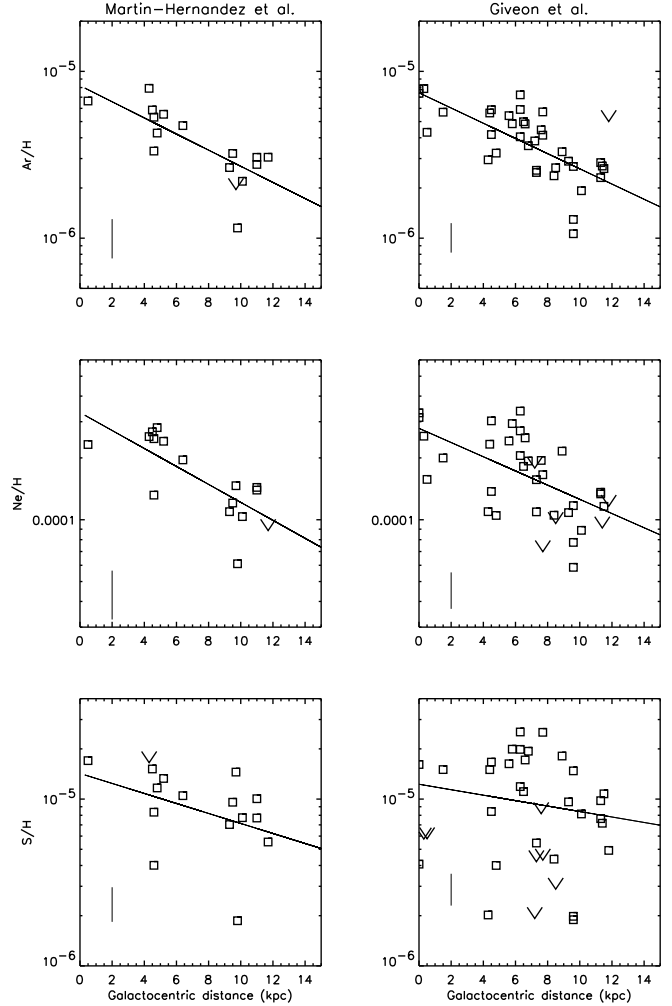}
\caption{Argon, neon, and sulfur abundances {\it vs.} galactocentric radius for the
full catalogs of Mart\'{\i}n-Hern\'{a}ndez et al. and Giveon et al. attenuation
and temperature corrected.
Solid lines are fits to the data, whose values are given in the fifth row of
Table \ref{gradlist}.}
\label{corr}
\end{figure}

Next, we apply extinction corrections to the full catalogs. 
Both Giveon et al. (2002)
and Mart\'{\i}n-Hern\'{a}ndez et al. (2002) 
employed the mid-infrared Draine (1989) extinction curve in their
analysis. For short wavelengths ($\lambda \le 7$ $\mu$m) both groups
assumed $A_\lambda / E(J-K) = 2.4 \lambda_{\mu {\rm m}}^{-1.75}$.
However, for long wavelengths ($\lambda > 7$ $\mu$m)
Giveon et al.  adopted the Draine mid-IR extinction curve
assuming a ``strong''
9.7 $\mu$m silicate absorption normalization, $A_{9.7}/E(J-K)=0.7$,
whereas  Mart\'{\i}n-Hern\'{a}ndez et al. assumed a ``weak''
normalization $A_{9.7}/E(J-K)=0.3$.  Here we adopt an
intermediate value, $A_{9.7}/E(J-K)=0.5$, for {\it both}
catalogs. The associated values of $A_\lambda/A_K$ for
the various emission lines are listed in Table 2.
The resulting gradients for the
attenuation-corrected full catalogs are given in the fourth row of Table
\ref{gradlist}.

\begin{table*}[]
\begin{center}
\caption{$A_\lambda/A_K$ line-extinction ratios.}
\vspace{0.5cm}
\begin{tabular}[h]{ccccccccc} \hline
Line              & Br$_\beta$ & Br$_\alpha$ & [ArII] & [ArIII] & [SIV] & [NeII] & [NeIII] & [SIII] \\  
Wavelength($\mu$m)&  2.60      &  4.05       &  7.0   &  9.0    & 10.5  & 12.8   &  15.5   & 18.7   \\
A$_\lambda$/A$_K$ &  0.72      &  0.33       &  0.13  &  0.64   &  0.70 & 0.28   & 0.22    & 0.32   \\
\hline
\end{tabular}
\label{extinc}
\end{center}
\end{table*}

We note that the attenuation-corrected argon 
and sulfur gradients are much
flatter than the uncorrected ones, while the neon
gradients are only slightly affected. This is mainly due to the large
extinction of \ariii~and \siii. These lines coincide
with the silicate absorption features at 9.7 and 18 $\mu$m 
(Draine \cite{draine}).

Next, we apply an electron temperature correction to the data. The temperature
correction accounts for possible variation in the \hii region temperatures
across the disk. 
Following Mart\'{\i}n-Hern\'{a}ndez et al. we assume that
the electron temperature varies as
\begin{equation}
T_{\rm e}=5000+5000\times{{R_{\rm gal}({\rm kpc})}\over{15}}.
\label{tempgrad}
\end{equation}
Applying this correction to the full catalogs results in the gradient given in
the fifth row of Table \ref{gradlist}.
We note that {\it both} the fine-structure lines and recombination lines are
temperature dependent. From a fit to the computations provided by Storey \&
Hummer (\cite{storey}), we find that the \bra\ recombination coefficient is
proportional to 
$T_{\rm e}^{-1.2}$. The fine-structure line emissivities are
proportional to $T_{\rm e}^{-0.5}$ (and to a temperature dependent exponential term).
Thus, the abundances vary as $T_{\rm e}^{-0.7}$. Given our assumed galactic
temperature profile (with temperaure increasing with radius) applying
the temperature correction steepens the resulting abundance gradients.
The attenuation and temperature
corrected data and gradients of the full catalogs are 
shown in Fig.~\ref{corr}.
The agreement between the two data sets is excellent.


We conclude that the Galactic metallicity gradients, as expressed by the
argon, neon, and sulfur abundances gradients, of Giveon et al. (\cite{giveon}) and
Mart\'{\i}n-Hern\'{a}ndez et al. (\cite{martin}) are consistent within the
error bars. 
This conclusion applies both to the
raw data and the fully corrected data.

\section*{Acknowledgments}

The ISO spectrometer data center at MPE is supported by DLR under
grants 50 QI 8610 8 and 50 QI 9402 3. Our research is supported by the
German-Israeli Foundation (grant I-0551-186.07/97).
We thank the referee for useful comments.

\end{document}